\documentclass{elsart3p}

\usepackage{epsfig}

\usepackage{amssymb}

\begin{document}

\begin{frontmatter}


\title{Impact parameter method calculations for fully differential ionization cross sections}

\author[author1]{F. J\'arai-Szab\'o} \ead{jferenc@phys.ubbcluj.ro} and
\author[author1]{L. Nagy}

\address[author1]{Faculty of Physics, Babe\c s-Bolyai University, str. Kogalniceanu 1, RO-400084 Cluj-Napoca, Romania}

\begin{abstract}
In this work our previous fully differential ionization cross section calculations using the semiclassical, impact parameter method are improved by a new method suitable to calculate impact parameter values corresponding to different momentum transfers. This goal is achieved by two successive steps. First, using the transverse momentum balance different projectile scattering angles are calculated for the binary and recoil peak regions as a function of the transferred momentum. Then, by treating the projectile scattering as a classical potential scattering problem, impact parameters are assigned to these scattering angles. The new method, which no longer contains empirical considerations, is tested calculating by fully differential ionization cross sections for single ionization of helium produced by fast C$^{6+}$ ions.
\end{abstract}

\end{frontmatter}


In the last 5 years there was an intensive effort to measure fully differential cross sections by kinematically
complete experiments. First, the group of Schulz reported interesting data for the
complete electron emission pattern in single ionization of helium by the impact of fast C$^{6+}$ ions for certain
momentum transfers \cite{Schulz2003,Schulz2004}. The three-dimensional images were obtained using experimentally
measured fully differential cross section values on a cold-target-recoil-ion-momentum spectrometer (COLTRIMS)
apparatus. Their results show a characteristic double-lobe structure with a binary peak and a smaller recoil peak.

On theoretical side several calculations exist being able to reproduce the experimental
data in the scattering plane (determined by the momentum of the scattered projectile and the
momentum transfer vectors) \cite{Madison2003,Foster2004a}.
At the same time, in the case of the perpendicular plane there is no agreement between the theoretical predictions and the experiment. Some authors have suggested that it may be important to include the internuclear interaction into the calculations \cite{Ciappina2006}. On the other hand, very recently, the importance of taking into account the uncertainties of the experimental measurements and performing a convolution of the theoretical results on the experimental resolution \cite{Fiol2006,Durr2007} was proved.

Previously, based on the semiclassical impact parameter method we constructed a theoretical model to calculate fully differential cross sections for single ionization of
helium by fast C$^{6+}$ ion impact \cite{Jarai2007}. Based on the semi-empirical version of the model a good agreement with the experiment was achieved in the scattering plane, while in the perpendicular plane a structure similar to that observed experimentally was obtained. It's main drawback was that the impact parameter values were selected based on the experimental data available in scattering plane.

In this work the model is improved by the introduction of a new method for calculating impact parameter values corresponding to different momentum transfers. Consequently, the empirical part of the description may be eliminated and the applicability of the semiclassical impact parameter method to calculate fully differential cross sections may be confirmed. Then, the method is tested in case of single ionization of helium produced by C$^{6+}$ ion projectile with an energy of $E_0 = 100$ MeV/u.


As described in \cite{Jarai2007} the semiclassical impact parameter method may be used to calculate fully differential ionization cross sections. In this approximation, the projectile is treated separately and it moves along classical trajectory. This implies that only the electron system needs to be described by a time-dependent Schr\"odinger equation, while the projectile follows classical laws throughout.

The ionization probability amplitudes for ionization of helium produced by fast C$^{6+}$ ions are calculated using first order time dependent perturbation theory. The initial state of the dielectronic system is described by a Hartree-Fock
wavefunction \cite{Clementi1974}, while the final state of the system is described by symmetric combination of a hydrogenic and a continuum radial wavefunction (calculated in the mean field of the final He$^+$ ion). Therefore, the ionization probability amplitude depending on the momentum transfer vector, ejected electron energy and electron ejection angles is reduced to a one-electron amplitude
\begin{equation}
a^{(1)} = -\frac{i\sqrt{2}}{v} \langle f_b | i_b \rangle
\int_{-\infty}^{+\infty}{dz\,e^{i \frac{E_f - E_i}{v}z} \langle f_c | V | i_b
\rangle}\,,
\end{equation}
where $i$ and $f$ represent the target system's initial and final electronic states, while the indices $b$ and $c$ represent bound and continuum states. Similarly, $E_i$ and $E_f$ are the energies of
the corresponding (unperturbed) states of the system and $V$ denotes the time-dependent interaction between projectile and active electron. The projectile velocity is denoted by $v$ and the integral is calculated through its classical trajectory along $z$ axis.

This amplitude is calculated expanding the final continuum-state wavefunctions into partial waves. As a result, amplitudes for transitions to ionized states with different angular momenta ($a_{l_fm_f}^{(1)}$) are obtained.

The fully differential cross sections relative to the momentum transfer value $q$, ejected electron energy $E$ and electron ejection angles $\theta$ and $\phi$ are obtained by the relation
\begin{equation}
\frac{d^5\sigma}{dE\;d\theta\;d\phi\;dq\;d\phi_q} = B\;\left| \sum_{l_f,m_f}
a_{l_fm_f}^{(1)} ({\mathbf B}) \right| ^2 \left|
\frac{dB}{dq}\right|,\label{fdcs1}
\end{equation}
where ${\mathbf B}$ is the impact parameter vector and $l_f$ and $m_f$ are quantum numbers of the partial waves describing the ejected electron.

Certainly, the most sensitive part of the model is to assign impact parameter values to a certain momentum transfer. This task may be completed in two successive steps.

First, the projectile scattering angle will be calculated by the use of the transverse momentum balance \cite{Ullrich1997} which states that the momentum transfer $q$ is the sum of the transverse components of electron's and residual ion's momenta. This vector relation may be written in scalar form as
\begin{equation}
p_{T\perp}^2 = p_{e\perp}^2 + q^2 - 2p_{e\perp}q\cos\phi\,,
\end{equation}
where $p_{T\perp}$ is the transverse momentum taken by the residual ion and $p_{e\perp} = p_e \sin \theta $ is the transverse momentum of the ionized electron. Further we assume, that the impact parameter is related to
the momentum transfer to the residual ion, and take into account the projectile-electron interaction separately.

Furthermore, let us now suppose that the momentum transfer $p_{T\perp}$ is modifying only the direction of the projectile momentum vector. Supposing that we have to deal with small scattering angles (valid for rapid collisions) and neglecting the terms being many times smaller than 1 the projectile scattering angle may be calculated as
\begin{equation}
\theta_{\rm proj} = \frac{\sqrt{p_e^2\sin^2\theta + q^2 - 2p_eq\sin\theta\cos\phi}}{p_0}\,,\label{scangle}
\end{equation}
where $p_0$ denotes the modulus of the projectile momentum.

In case of binary peak, one has to deal with $\phi = 0$ while in case of recoil
peak the value of this angle is 180$^o$. In both cases the maxima are at $\theta=90^o$.
For these peaks the scattering angles are
\begin{equation}
\theta_{\rm proj}^{\rm binary} = \frac{\left|p_e - q \right|}{p_0}\;\;\;{\rm and}\;\;\;\;\;
\theta_{\rm proj}^{\rm recoil} = \frac{\left|p_e + q \right|}{p_0}\,.
\end{equation}
For the experimentally studied $q=0.75$ a.u. and $E=6.5$ eV case these angles have values of $\theta_{\rm proj}^{\rm binary} = 4.2073\cdot 10^{-8}$ rad and $\theta_{\rm proj}^{\rm recoil} = 1.0337\cdot 10^{-6}$ rad.

Second, our next task is to assign impact parameter values to the projectile scattering angles. Accordingly, the projectile scattering will be treated as a classical potential scattering problem in the field of the target helium system \cite{Newton2002}. Therefore, the scattering angle may be calculated as
\begin{equation}
\theta_{\rm proj} = \pi -2B\int_{r_0}^{\infty}\frac{dr}{r^2}\left( 1 - \frac{U(r)}{E_0} - \frac{B^2}{r^2}\right)^{-1/2}\,,\label{csint}
\end{equation}
where $r_0$ is the distance of closest approach defined by $\left( 1 - \frac{U(r)}{E_0} - \frac{B^2}{r^2}\right)=0$ and $U(r)$ is the scattering potential. The simplest way to include the effect of the electrons around the target nucleus is to consider the potential to be a product of the Coulomb potential and the Bohr-type screening function \cite{Everhart1955}
\begin{equation}
U(r) = \frac{Z_{\rm proj} Z_{\rm target}}{r} e^{-\frac{r}{a}}\,.
\end{equation}
If, particularly, the Thomas-Fermi potential \cite{Lindhard1968} is used, then for the C$^{6+}$ + He colliding system the value of parameter  $a$ will be given by
\begin{equation}
a=\frac{0.8853}{Z_{\rm target}^{0.23}}
\end{equation}
expressed in a.u.

\begin{figure}
\begin{center}
\epsfxsize=7.5cm
\epsfbox{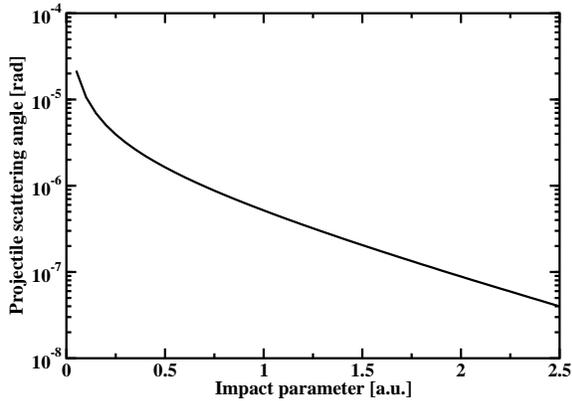}
\end{center}
\caption{\label{ipB}Calculated projectile scattering angles for different impact parameter values.}
\end{figure}

Using this potential the integral (\ref{csint}) may be calculated numerically. Figure \ref{ipB} shows the results of our calculations, namely the dependence of the scattering angle on impact parameter. From this data it turns out that impact parameters corresponding to previously calculated binary and recoil scattering angles are 2.47 and 0.68 a.u., respectively.

\begin{figure}
\begin{center}
\epsfxsize=7.5cm
\epsfbox{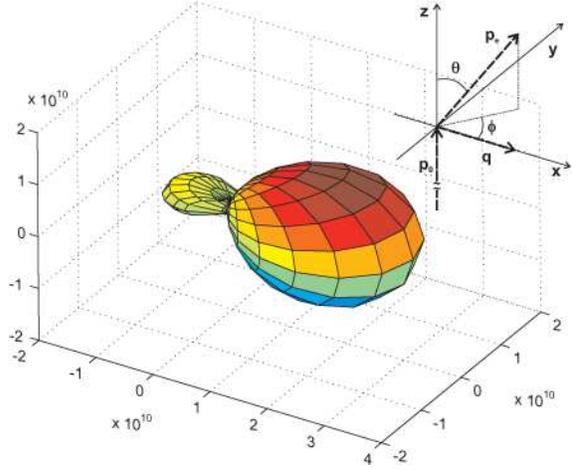}
\end{center}
\caption{\label{3d1}3D image of the electron emission
pattern for single ionization of helium produced by 100 MeV/u C$^{6+}$
projectile impact calculated by semiclassical impact parameter method.
The ejected electron energy is $E_e = 6.5$ eV and the
momentum transfer is $q=0.75$ a.u.}
\end{figure}

Just for visualization purposes, based on our new theoretical results we have constructed the 3D image of the electron emission pattern for the studied case. The results presented on Figure \ref{3d1} are calculated using the previously determined impact parameter values. The transition between these impact parameters is realized smoothly in the $0<\theta<50^o$ and $130^o<\theta<180^o$, $90^o<\phi<270^o$ transition regions. As sketched on the figure, the initial projectile direction is along the $z$ axis and the momentum transfer vector $\mathbf q$ is pointing nearly in $x$ direction. In the 3D graph one can observe the presence of the characteristic double-lobe structure towards the x axis with binary peak at $\theta = 90^o$, $\phi = 0^o$ and recoil peak at $\theta = 90^o$, $\phi = 180^o$. The electron emission pattern is almost identical with our previous result obtained by the semi-empirical model \cite{Jarai2007}.

\begin{figure}
\begin{center}
\epsfxsize=7.5cm
\epsfbox{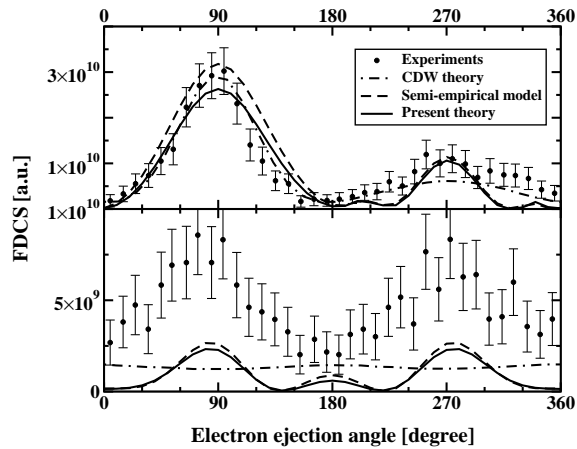}
\end{center}
\caption{\label{0.75planes}Theoretical results in scattering (top) and
perpendicular (bottom) planes compared with experiments \cite{Fiol2006} for the
same case as in figure \ref{3d1}. The solid curve shows the present theory,
dashed line represent results of the semi-empirical version of the model \cite{Jarai2007}, while the dash-dotted line is obtained by the CDW model \cite{Fiol2006}.}
\end{figure}

In order to analyze in detail the obtained results, cross section values for scattering plane and perpendicular plane are plotted separately in figure \ref{0.75planes}. Thus, the results of the present theory may be compared in absolute value to experimental data, previous results of the semi-empirical version of the model and CDW calculations.

The top panel of the figure shows fully differential cross sections in the scattering plane for different electron ejection angles $\phi$. The present model gives good agreement with experimental data of Schulz et al. \cite{Schulz2003}. However, a small difference may be detected between the current version and the semi-empirical version of the model mainly in the binary peak region. The difference may be explained knowing that in case of the previous semi-empirical model impact parameters of 2.2 and 0.7 a.u. have been selected based on the available experimental data \cite{Jarai2007}. Even so, the difference between results looks to be smaller than the experimental errors.

In case of the perpendicular plane almost the same results as the semi-empirical ones have been obtained. The bottom
panel of figure \ref{0.75planes} shows fully differential cross sections in this plane for different electron ejection angles $\theta$. The curve shows the same behavior as the
experimental data with strong maxima at $\theta = 80^o$ and $\theta = 280^o$. A
third smaller maximum is also obtained at direction of $\theta = 180^o$.

Here it has to be noted that, compared to the isotropic results of
the CDW model,  a better agreement in shape has been obtained  by the
present metod. However, the magnitude of the cross section
is smaller than the experimental one. The recently
reported inclusion of the experimental momentum uncertainties \cite{Fiol2006}
should also improve the agreement between theory and experiments by increasing
the cross sections in the perpendicular plane. Our result is consistent with
the conclusions of D\"urr et al. \cite{Durr2007}, that the experimental
uncertainties are responsible only partly for the structure observed in the
perpendicular plane, half of the value of the maxima may be due to some real
physical effect.

In conclusion, the theoretical model based on the first order, semiclassical, impact parameter approximation for describing kinematically complete experiments has been revised by eliminating any empirical considerations from it. This has been achieved by calculating impact parameter values corresponding to different momentum transfers. Then, the new model was applied for studying single ionization of helium by impact with fast C$^{6+}$ ions. As its predecessor, the model describes well the fully differential cross sections for relatively small momentum transfer values. The characteristic structures in the perpendicular plane have also been reproduced, discrepancies with the experiments are only in the magnitude of the cross sections, which may be explained by the experimental uncertainties.

Finally, it has to be mentioned, that the main goal of this work was not to improve the previous results of the semiclassical impact parameter model used to calculate fully differential cross sections, but to create a pure theoretical background for it. Accordingly, all previous results of the model have been confirmed, and we have  given a theoretical method for obtaining the impact parameter used in the calculations.

The present work has been supported by the Romanian National Plan for Research (PN II),
contract No. ID 539.


\begin{thebibliography}{00}

\bibitem{Schulz2003}
M. Schulz \etal, Nature 422 (2003) 48 and references therein.
\bibitem{Schulz2004}
M. Schulz, R. Moshammer, D. Fischer and J. Ullrich, J. Phys. B: At. Mol. Opt. Phys. 37 (2004) 4055.
\bibitem{Madison2003}
D. H. Madison \etal Phys. Rev. Lett. 91 (2003) 253201.
\bibitem{Foster2004a}
M. Foster \etal, J. Phys. B: At. Mol. Opt. Phys. 37 (2004) 1565.
\bibitem{Ciappina2006}
M. F. Ciappina and W. R. Cravero, J. Phys. B: At. Mol. Opt. Phys. 39 (2006) 2183.
\bibitem{Fiol2006}
J. Fiol, S. Otrantoand, R. E. Olson, J. Phys. B: At. Mol. Opt. Phys. 39 (2006) L285.
\bibitem{Durr2007}
M. D\"urr \etal, Phys. Rev. A 75 (2007) 062708.
\bibitem{Jarai2007}
F. J\'arai-Szab\'o, L. Nagy, J. Phys. B: At. Mol. Opt. Phys. 40 (2007) 4259.
\bibitem{Clementi1974}
E. Clementi and C. Roetti, At. Data Nucl. Data Tables 14 (1974) 177.
\bibitem{Ullrich1997}
J. Ullrich \etal J. Phys. B: At. Mol. Opt. Phys. 30 (1997) 2917.
\bibitem{Newton2002}
Roger G. Newton, Scattering Theory of Waves and Particles, Second edition, Mineola, New York (2002) ISBN 0-486-42535-5.
\bibitem{Everhart1955}
E. Everhart, G. Stone and R. J. Carbone, Phys. Rev. 99 (1955) 1287.
\bibitem{Lindhard1968}
J. Lindhard, V. Nielsen and M. Scharff, K. Dan. Vidensk. Selsk. Mat. Fys. Medd. 36 (1968) 10.

\end{thebibliography}
\end{document}